\documentclass[aps,prd]{revtex4}

\usepackage{graphicx}
\usepackage{lscape}
\usepackage{indentfirst}
\usepackage{latexsym}
\usepackage{multirow}
\usepackage{tabls}
\usepackage{epsfig}
\usepackage{color}
\usepackage{amssymb}

\usepackage{amsfonts}
\usepackage{amsmath}
\usepackage{bm}
\usepackage{mathrsfs}

\def\dprime{ {\prime \prime} }

\def\b{ \beta }

\def\cB{ {\cal B} }

\def\cE{ {\cal E} }

\def\cJ{ {\cal J} }

\def\b0{ {\bf 0} }

\def\lsim{\mathrel{\rlap{\lower3pt\hbox{\hskip1pt$\sim$}}
    \raise1pt\hbox{$<$}}}                
\def\gsim{\mathrel{\rlap{\lower3pt\hbox{\hskip1pt$\sim$}}
    \raise1pt\hbox{$>$}}}         

\def\coordeq{ \, \mathrel{ \rlap{\hbox{\hskip-2.5pt$=$} }
    \raise4pt\hbox{$\cdot$}} \, }                

\begin{document}

\title{Radiation reaction at 3.5 post-Newtonian order in effective field theory}

\def\addJPL{Jet Propulsion Laboratory, California Institute of Technology, Pasadena, California 91109, USA}
\def\addCaltech{Theoretical Astrophysics, California Institute of Technology, Pasadena, California 91125, USA}
\def\addUPitt{Pittsburgh Particle physics Astrophysics and Cosmology Center (PITT PACC)\\Department of Physics and Astronomy, University of Pittsburgh, Pittsburgh, Pennsylvania 15260, USA}

\author{Chad R. Galley} 
\affiliation{\addJPL}
\affiliation{\addCaltech} 

\author{Adam K. Leibovich}
\affiliation{\addUPitt}

\date{\today}

\begin{abstract}
We derive the radiation reaction forces on a compact binary inspiral through 3.5 order in the post-Newtonian expansion using the effective field theory approach. We utilize a recent formulation of Hamilton's variational principle that rigorously extends the usual Lagrangian and Hamiltonian formalisms to dissipative systems, including the inspiral of a compact binary from the emission of gravitational waves. We find agreement with previous results, which thus provides a non-trivial confirmation of the extended variational principle. The results from this work nearly complete the equations of motion for the generic inspiral of a compact binary with spinning constituents through 3.5 post-Newtonian order, as derived entirely with effective field theory, with only the spin-orbit corrections to the potential at 3.5 post-Newtonian remaining.
\end{abstract}

\maketitle

\section{Introduction}

The advent of advanced ground-based gravitational wave (GW) interferometer detectors (i.e., advanced LIGO and advanced VIRGO)  brings an increasing demand for more accurate waveform templates to be used for detecting gravitational waves and for extracting information about the parameters associated with a source, such as the masses, spins, distance, and sky location. Currently, a goal of the GW source-modeling community is to produce inspiral waveforms accurate through at least 3.5 post-Newtonian (PN) order. The PN expansion is a perturbation theory for the gravitational field and the binary's motion in the weak-field and slow-motion limits (see \cite{Blanchet:LRR} for a review). The equations for the relative motion of the binary are known already through 3.5PN order, even when including the spin angular momenta of the binary's constituents (see, e.g., \cite{Blanchet:LRR, Porto:PRD73, PortoRothstein:PRL97, PortoRothstein:PRD78_2, Hartung:2011te} and references therein). However, the gravitational wave flux and the waveform, especially, are not yet known to such a high order for spinning binary inspiral sources. 

High-accuracy waveforms and source-modeling are also important for matching post-Newtonian inspiral waveforms to numerical simulations of binary mergers for purposes of parameter estimation (see e.g., \cite{MacDonald:2011ne}) and for accurately calibrating phenomenological models like Effective One Body \cite{BuonannoDamour:PRD59} and hybrid models \cite{Ajithetal:CQG24, Ajithetal:PRD77, Ajith:2009bn, Santamaria:2010yb}. These phenomenological models may be used to construct relatively cheap template banks without having to run a prohibitively large number of expensive numerical simulations of binary mergers. This is especially true if combined with an efficient template bank compression and representation scheme such as provided by the Reduced Basis method \cite{RB:PRL}.

The Effective Field Theory (EFT) approach \cite{GoldbergerRothstein:PRD73} offers an efficient and algorithmic computational framework compared to traditional methods \cite{Blanchet:LRR, FutamaseItoh:LRR} and has rapidly made useful contributions towards the  goal of 3.5PN-accurate inspiral waveforms. To date, this includes the calculation of the PN corrections to the binding potential, including spin angular momenta of the binary's component masses, through 3PN order \cite{GoldbergerRothstein:PRD73, KolSmolkin:CQG25, GilmoreRoss:PRD78, FoffaSturani:PRD84, Porto:PRD73, Perrodin:2010dy, Porto:CQG27, Levi:PRD82_1, PortoRothstein:PRL97, PortoRothstein:0712.2032, PortoRothstein:PRD78}, the spin1-spin2 terms computed at 4PN \cite{Levi:PRD85}, the leading order radiation reaction force at 2.5PN \cite{GalleyTiglio:PRD79}, the multipole moments needed for calculating the gravitational wave flux through 3PN \cite{PortoRossRothstein:JCAP1103}, and the moments needed for calculating the waveform amplitude corrections through 2.5PN \cite{Porto:2012as}.

In this paper, we calculate the radiation reaction forces that appear at 3.5PN order (spin effects do not enter radiation reaction until 4PN). In doing so, we nearly complete the PN equations of motion for the generic inspiral of a compact binary with spinning constituents as computed {\it entirely} in the EFT framework. Only the spin-orbit correction to the potential at 3.5PN order is remaining to be computed in EFT. This is a rather remarkable achievement given that the EFT approach was introduced almost eight years ago \cite{GoldbergerRothstein:PRD73}.

Computing radiative effects in the EFT approach presents a unique challenge since the formalism makes heavy use of an action formulation of the binary system. More specifically, it is well-known that Lagrangians and Hamiltonians are not generally applicable to dissipative systems, which would
make computing the PN radiation reaction forces in EFT very difficult. 
In \cite{GalleyTiglio:PRD79}, it was indicated how this might be overcome using a language and notation from quantum field theory but was not given a rigorous foundation within a purely classical mechanical context. Nevertheless, as a demonstration of the formalism, the 2.5PN radiation reaction force and the gravitational waveform from the binary's leading order quadrupole moment were computed \cite{GalleyTiglio:PRD79} and shown to agree with existing results \cite{BurkeThorne:Relativity, Burke:JMathPhys12}, thus lending credibility to the method. Recently, one of us (CRG) gave a rigorous extension of Hamilton's variational principle in \cite{Galley:dissCM} that yields a Lagrangian (and Hamiltonian) formulation that suitably and correctly describes generally dissipative systems. We use this formalism here, together with EFT, to compute the 3.5PN radiation reaction force and find agreement with previously published results \cite{IyerWill:PRL70, IyerWIll:PRD52}. This agreement lends non-trivial confirmation for the validity of the extended variational principle for dissipative systems described in \cite{Galley:dissCM} (see also the examples given in that reference).

This paper is organized as follows. Section \ref{sec:overview} gives an overview of the formalisms needed to compute the radiation reaction force at 3.5PN order in EFT. Specifically, Section \ref{sec:eft} reviews the EFT of compact binary inspirals and Section \ref{sec:dissCM} reviews the recently formulated extension of Hamilton's variational principle for {\it dissipative} systems \cite{Galley:dissCM}. Section \ref{sec:radrxn35} discusses the computation of the 3.5PN radiation reaction force in EFT where agreement is shown with previous results \cite{IyerWill:PRL70, IyerWIll:PRD52}. Section \ref{sec:conclusion} concludes the paper and Appendix \ref{appendixA} outlines in detail the EFT calculation of the leading order 2.5PN radiation reaction force of Burke and Thorne \cite{BurkeThorne:Relativity, Burke:JMathPhys12}.

\section{Overview of effective field theory and dissipative mechanics}
\label{sec:overview}

We begin by reviewing the effective field theory of compact binary inspirals, focusing mainly on the radiation sector, and end the section by reviewing how dissipative (e.g., radiative) effects can be handled within the newly developed mechanics for dissipative systems \cite{Galley:dissCM}.

\subsection{Effective field theory of compact binary inspirals}
\label{sec:eft}

The EFT approach, introduced by Goldberger and Rothstein in \cite{GoldbergerRothstein:PRD73}, separately treats the relevant scales of the binary by successively ``integrating out'' the smaller scales thereby yielding a hierarchy of EFTs that are related to each other through so-called matching calculations. The effects of short-distance physics in a large-distance effective theory are parameterized in a manner consistent with the symmetries (e.g., general coordinate invariance) as discussed in more detail below. 

The slow inspiral of compact binaries has three relevant scales, from smallest to largest: the size of the compact objects (COs) $R_m$, their orbital separation $r$, and the wavelength of the emitted gravitational waves $\lambda$. The first EFT describes the extended masses in the point particle approximation. To incorporate the finite size of the CO one appends to the point particle action all possible interaction terms that are consistent with general coordinate invariance and reparameterizations of the worldline.
This leads to a worldline EFT, for one of the COs (e.g., spherical and non-spinning), that is described by the action \cite{GoldbergerRothstein:PRD73}
\begin{equation}
	S_{\rm CO} =  -m \int d\tau + C_E \int d\tau \, {\cal E}_{\alpha \beta} {\cal E}^{\alpha \beta} + C_B \int d\tau\, {\cal B}_{\alpha \beta} {\cal B}^{\alpha \beta} + \cdots 
	\label{ppaction}
\end{equation}
where ${\cal E}_{\alpha\beta}$ and ${\cal B}_{\alpha \beta}$ are the electric and  magnetic parts of the Weyl curvature tensor.
The coefficients $C_E, C_B, \ldots$ are determined via a matching calculation, wherein a chosen quantity is calculated in the effective theory and in the long-distance limit of the ``full'' theory for the CO. One can show that these extra terms in the action contribute to the binding potentials (due to induced quadrupole moments) starting at 5PN for non-spinning COs.  We can therefore ignore these terms for the work presented here.

As one ``zooms out'' from $R_m$ to the orbital radius $r$, the system is described by General Relativity coupled to two COs that are each described by the worldline EFT in (\ref{ppaction}). At this scale, the particles interact with two kinds of gravitational perturbations. The first describes the nearly instantaneous potentials that bind the two particles in orbit (for further details see \cite{GoldbergerRothstein:PRD73}).
The second is the long-wavelength GWs emitted by the binary.

As one ``zooms out'' from $r$, the binary itself is described in the point particle approximation by a composite object \cite{GoldbergerRothstein:PRD73, GalleyTiglio:PRD79, GoldbergerRoss:PRD81}. This composite object radiates gravitational waves from its time-dependent multipole moments that can be calculated by matching onto the radiative moments of the binary at the orbital scale. The worldline EFT for this radiating object has an action given by \cite{GoldbergerRoss:PRD81}
\begin{align}
	S_{\rm rad} [ x^\mu, h_{\mu\nu} ] = {} & - \int d\tau \, M(\tau) - \frac{1}{2} \int d\tau \, L_{ab} (\tau) \big[ \Omega_L^{ab} + u^\mu \omega_\mu {}^{ab} (\tau) \big] + \frac{1}{2} \sum_{n=0}^\infty \int d\tau \, c_n^{(I)} I^{ab a_1 \cdots a_n } (\tau) \nabla_{a_1} \cdots \nabla_{a_n} {\cal E}_{ab} (x^\mu)  \nonumber \\
	&  + \frac{1}{2} \sum_{n=0}^\infty \int d\tau \, c_n^{(J)} J^{ab a_1 \cdots a_n} (\tau) \nabla_{a_1} \cdots \nabla_{a_n} {\cal B}_{ab} (x^\mu) + \cdots
\label{radEFT1}
\end{align}
where $x^\mu (\tau)$ are the worldline coordinates ($\tau$ beings its proper time), $I^{ab a_1 \cdots a_n}(\tau)$ and $J^{ab a_1 \cdots a_n}(\tau)$ are the symmetric, trace-free (STF) mass and current multipole moments, respectively, of the composite object \cite{GoldbergerRoss:PRD81, Ross:2012fc}, $\Omega_L^{ab}$ is the angular frequency of the body's rotation as measured in a locally flat Lorentz frame, and $\omega_\mu{}^{ab}$ are the spin connection coefficients, which couple to the total angular momentum $L^{ab} = - L^{ab}$. Also, lower case roman letters takes values in $\{1,2,3 \}$. The mass of the body is taken to be generally time-dependent $M(\tau)$ as the body may lose rest-mass energy, as measured by a distant observer, via gravitational wave emission. A local Lorentz frame is attached to the worldline such that $e_0^\mu(\tau) = u^\mu (\tau) = dx^\mu / d\tau$ and $e^\mu_a(\tau)$ are space-like vectors that rotate with the body and thus account for its spin dynamics. The first several $c_n^{(I, J)}$ coefficients are conventionally taken to be
\begin{align}
	c_0^{(I)} = 1, \qquad c_0^{(J)} = - \frac{4}{3} , \qquad c_1^{(I)} = \frac{1}{3} .
\end{align}

It is important to note that (\ref{radEFT1}) describes {\it any} multipolar, extended body that has a size smaller than the wavelength of gravitational waves it emits. Therefore, (\ref{radEFT1}) is a rather general and model-independent description of such a system. However, using the multipole moments computed in the PN expansion for compact binary inspirals \cite{PortoRossRothstein:JCAP1103, Ross:2012fc} one may use (\ref{radEFT1}) to study radiative effects, such as radiation reaction, in compact binary systems.

As with any field theory coupled to point particles, including the EFTs reviewed here, divergences will appear. However, there are an infinite number of parameters in the theory (e.g., $c_n^{(I)}$, $c_n^{(J)}$) so that divergences can {\it always} be absorbed into these coupling constants. Interestingly, these parameters can exhibit a {\it classical} renormalization group flow due to gravitational screening effects, which manifest as logarithmic divergences in the potentials. Unlike with traditional approaches for the PN expansion \cite{Blanchet:LRR}, divergences in the EFT approach have a natural place and interpretation in the context of renormalization group theory \cite{GoldbergerRothstein:PRD73}. Spin angular momenta for the binary's component masses can be included as in \cite{Porto:PRD73}.

\subsection{Classical mechanics for dissipative systems}
\label{sec:dissCM}

Determining the evolution of the compact binary in EFT is achieved by integrating out the gravitational perturbations from the action. In practice, this is accomplished by solving for the gravitational perturbations and substituting these solutions into the original action to yield an {\it effective action} for the worldlines of the component masses \cite{GoldbergerRothstein:PRD73, KolSmolkin:PRD77}, which is efficiently carried out using Feynman diagrams \cite{GoldbergerRothstein:PRD73}. This procedure is well-suited for conservative interactions, such as for computing PN corrections to the binding potential of a compact binary, but requires modification when studying dissipative systems, such as the inspiral of a compact binary due to the emission of gravitational radiation, the reason being that dissipative systems generally do not admit Lagrangian or Hamiltonian descriptions.


Recently, one of us (CRG) introduced a rigorous variational calculus for Hamilton's principle of stationary action that includes systems exhibiting {\it dissipative} effects and thus generalizes the usual action principle \cite{Galley:dissCM}. The derivation and the details will not be given here but instead refer the reader to \cite{Galley:dissCM}. However, we will review the relevant formalism needed specifically for computing the radiation reaction force in this paper. 

The total system formed by the gravitational perturbations $h_{\mu\nu}$ and the worldlines of the compact bodies in a binary system, $\vec{x}_K(t)$ for $K=1, 2$, is closed. Only when the gravitational perturbations are integrated out is the dynamics of the worldlines open. When integrating out the long wavelength gravitational perturbations at the level of the action one must be careful when applying Hamilton's principle of stationary action to the effective action since it is formulated by specifying {\it boundary} conditions in time, not {\it initial} conditions. This observation may seem innocuous except that the resulting effective action for the worldlines describes conservative (i.e., non-dissipative) dynamics -- the Green function for the gravitational perturbations that appear in the effective action are time-symmetric (as these are the ones satisfying boundary conditions according to Sturm-Liouville theory) and hence do not account for the dissipative effects of radiation reaction \cite{GalleyTiglio:PRD79} . 

To overcome this problem, one formally doubles the degrees of freedom \cite{Galley:dissCM} so that $h_{\mu\nu} \to ( h_{\mu\nu 1} , h_{\mu\nu 2})$ and $\vec{x}_K \to ( \vec{x}_{K 1} , \vec{x}_{K 2} )$ then constructs the following action,
\begin{align}
	S [ \vec{x}_{K1}, \vec{x}_{K2}, h_{\mu\nu 1}, h_{\mu\nu 2} ] = S [ \vec{x}_{K 1}, h_{\mu\nu 1} ] - S [ \vec{x}_{K 2}, h_{\mu\nu 2} ]
\end{align}
where each action on the right side consists of the Einstein-Hilbert action (here, gauge-fixed in the Lorenz gauge and expanded around flat spacetime) and the worldline EFT action (\ref{radEFT1}). Integrating out the long wavelength gravitational perturbations using Feynman diagrams at the desired PN order (see \cite{GalleyTiglio:PRD79} for the Feynman rules of the radiation EFT and how to construct the Feynman diagrams) gives the effective action for the open dynamics of the binary's inspiral. After all variations are performed one is free to identify the doubled variables with the physical ones so that, for example, $\vec{x}_{K 1}, \vec{x}_{K 2} \to \vec{x}_K$. This limit will be called the {\it physical limit}.


It is convenient, but not necessary, to change variables from the $(1,2)$ variables to $\pm$ variables where 
\begin{align}
	\vec{x}_{K+} & = \frac{ \vec{x}_{K 1} + \vec{x}_{K 2} }{ 2 } \\
	\vec{x}_{K-} & = \vec{x}_{K 1} - \vec{x}_{K 2}
\end{align}
This change of variable is motivated by the physical limit since then $\vec{x}_{K-} \to 0$ and $\vec{x}_{K+} \to \vec{x}_K$ where $\vec{x}_K$ is the physical worldline of the $K^{\rm th}$ body. By computing the following variation \cite{Galley:dissCM},
\begin{align}
	0 = \frac{ \delta S_{\rm eff} [ \vec{x}_{1 \pm}, \vec{x}_{2 \pm} ] }{ \delta \vec{x}_{K-} (t) } \bigg|_{ \substack{ \vec{x}_{K-} = 0 \\ \vec{x}_{K+} = \vec{x}_K } }
\label{eq:vard1}
\end{align}
one obtains a set of worldline equations of motion that properly incorporates radiation reaction effects \cite{Galley:dissCM, GalleyTiglio:PRD79}. 
It is important to note that (\ref{eq:vard1}) receives non-zero contributions from terms in the effective action that are perturbatively {\it linear} in $\vec{x}_{K-}$ or its time derivatives. Therefore, all terms of quadratic order or higher in any ``$-$'' variables can be dropped from the effective action for the purposes of deriving the worldline equations of motion. We will take advantage of this property in the course of the calculations below in Section \ref{sec:radrxn35} and Appendix \ref{appendixA}.

In the remaining sections, we compute the terms of the effective action at 3.5PN order by computing the corresponding Feynman diagrams in the extended action formalism of \cite{Galley:dissCM}. Once the effective action is derived in Section \ref{sec:radrxn35}, we then apply the variational principle in (\ref{eq:vard1}) to obtain the radiation reaction forces on the binary at 3.5PN order.

\section{Radiation reaction through 3.5PN}
\label{sec:radrxn35}

In this section we lay out how to calculate the radiation reaction at 3.5PN in the EFT.  
However, the first non-conservative term in the acceleration enters at 2.5PN order and is due to radiation reaction. We include in Appendix \ref{appendixA} a  detailed calculation of the radiation reaction at 2.5PN using the EFT.  The 3.5PN terms can be derived similarly so we will often refer the reader to Appendix \ref{appendixA} for formulae.

The relative acceleration $a^i = a_1^i - a_2^i$ for a non-spinning two-body system can be expanded as
\begin{equation}
a^i = a_0^i +\epsilon a_1^i + \epsilon^2 a_2^i + \epsilon^{2.5} a_{2.5}^i + \epsilon^3 a_{3}^i + \epsilon^{3.5} a_{3.5}^i + \cdots 
\label{eq:pnacc1}
\end{equation}
where $\epsilon = 1$ counts the post-Newtonian order.  The motion through 2PN order is conservative, meaning that the motion can be characterized by the conserved total energy and angular momentum.  Dissipative effects first arise at order 2.5PN due to gravitational radiation reaction.  The 3PN terms are again conservative while the 3.5PN terms are the first post-Newtonian corrections to radiation reaction.  

The first term in (\ref{eq:pnacc1}) is just the Newtonian acceleration,
\begin{equation}
a_0^i = -\frac{m}{r^2}n^i,
\label{newta}
\end{equation}
while the second term is the 1PN correction, derived from the Einstein-Infeld-Hoffmann (EIH) Lagrangian \cite{EinsteinInfeldHoffmann:AnnMath39},
\begin{align}
a_1^i &=-(3+\eta) \frac{m}r a^i - m\eta\frac{ \vec{a} \cdot \vec{x}}{r^2}n^i - \frac{1-3\eta}2 v^2 a^i - (1-3\eta) \vec{v} \cdot \vec{a} \, v^i -\frac{m}{r^2}\left\{n^i\left[-\frac{m}{r} + \frac32(1+\eta)v^2 - \frac32 \eta \dot r^2\right] - v^i \dot r (3+\eta)\right\}
\label{1pnnotreduced} \\
&= -\frac{m}{r^2}\left\{n^i\left[-2(2 + \eta)\frac{m}r + (1 + 3\eta)v^2 - \frac32 \eta\dot r^2\right] - 2(2-\eta) \dot{r} v^i\right\},
\label{1pn}
\end{align}
where $m$ is the total mass, $\eta = m_1 m_2/m^2$ is the symmetric mass ratio, $n^i = x^i/r$ with $x^i = x_1^i - x_2^i$ the separation between the point masses, and $v^i = dx^i / dt$. The EIH Lagrangian was calculated using the EFT in~\cite{GoldbergerRothstein:PRD73, KolSmolkin:CQG25}.  To go from Eq.~(\ref{1pnnotreduced}) to Eq.~(\ref{1pn}) it is necessary to order-reduce by substituting the ${\cal O}(\epsilon^0)$ expression for the acceleration. By including the acceleration at ${\cal O}(\epsilon^{2.5})$ when performing the order-reduction, we get a contribution to the 3.5PN acceleration.  This will be discussed in more detail below.
In the rest of this section we derive the radiation reaction through 3.5PN using the EFT.

\subsection{Feynman diagrams}
\label{sec:feynman}

\begin{figure}
	\center
	\includegraphics[width=4cm]{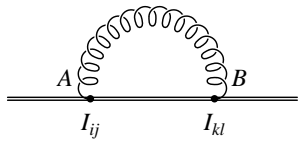}
	\caption{The Feynman diagram giving the contribution to the radiation reaction from the mass quadrupole.  $A$ and $B$  label worldlines that have been doubled in the extended variational principle.}
	\label{fig:massquad}
\end{figure}

To calculate the radiation reaction, we begin from the worldline action (\ref{radEFT1}) and integrate out the electric and magnetic parts of the Weyl curvature tensor giving, to the order we require, the effective action 
\begin{equation}
S^{\rm 3.5PN}_{\rm eff} = S_{mq} + S_{mo} + S_{cq},
\label{Seff}
\end{equation}
where the terms on the right-hand side stand for mass quadrupole ($mq$), mass octupole ($mo$), and current quadrupole ($cq$), respectively.
As discussed in Appendix~\ref{appendixA}, at 2.5PN this is accomplished by calculating the diagram in Fig.~\ref{fig:massquad}.  At 2.5PN, each vertex is given by the leading order mass quadrupole moment $I^{ij} = I_0^{ij} + {\cal O}(\epsilon)$ where
\begin{equation}
I^{ij}_{0} = \sum_K m_K \left(x_K^i x_K^j - \frac13\delta^{ij} x^2_K \right),
\label{massquadLO}
\end{equation}
and where $K$ labels the worldlines in the binary $K=1,2$ and the subscript on $I^{ij}$ denotes the (relative) PN order in $\epsilon$. To calculate at 3.5PN, we need to include PN corrections to the mass quadrupole,
\begin{align}
I^{ij} &= I^{ij}_{0} + \epsilon I^{ij}_{1} + {\cal O}(\epsilon^{1.5}) \nonumber\\
  &= 
  	\sum_K m_K \bigg\{ x_K^i  x_K^j  + \epsilon \bigg[\bigg(\frac32 v_K^2 - \sum_{L\neq K}\frac{G m_L}r  \bigg)  x_K^i x_K^j  
		+ \frac{11}{42}\frac{d^2}{dt^2} \big(x_K^2 x_K^i x_K^j \big)  -\frac43\frac{d}{dt} \big( \vec{x}_K\cdot \vec{v}_K x_K^i x_K^j\big) \bigg] \bigg\}_{\rm TF} + {\cal O}(\epsilon^{1.5})
\label{massquad1PN}\\
  &= \bigg\{\mu x^i x^j + \epsilon \mu \bigg[\bigg(\frac{29}{42}(1-3\eta)v^2 - \frac17(5-8\eta)\frac{m}r\bigg) x^i x^j + \frac1{21}(1-3\eta) \big(11 r^2 v^i v^j - 12 r \dot r x^{(i} v^{j)}\big) \bigg]\bigg\}_{\rm TF} + {\cal O}(\epsilon^{1.5}) ,
\label{massquad1PNcom}
\end{align}
where TF denotes taking the trace-free part of the enclosed expressions and in the last line we have gone to center-of-mass coordinates and $\mu = \eta m$ is the reduced mass.  However, the diagram in Fig.~\ref{fig:massquad} is the same whether one includes the 1PN corrections to the mass quadrupole or not, so we get the effective action (\ref{eq:effaction25c}) derived in Appendix \ref{appendixA}, namely,
\begin{equation}
S_{mq} = \frac{G}5 \int dt \, I_-^{ij}(t)I_{ij+}^{(5)}(t).
\end{equation}
where $I_-^{ij} \equiv I^{ij}_1 - I^{ij}_2$ and $I_+^{ij} \equiv ( I_1^{ij} + I_2^{ij} )/2$, and $I_A^{ij}$ for $A=1,2$ is the mass quadrupole moment formed by the $A^{\rm th}$ worldline ($A$ labels the doubled worldlines, not the components of the two-body system, which is indexed by $K=1,2$).
By expanding in $\epsilon$, 
\begin{equation}
S_{mq} = \frac{G}5\left[\epsilon^0 \int dt I_{0-}^{ij}(t)I_{0ij+}^{(5)}(t) + \epsilon \int dt I_{0-}^{ij}(t)I_{1ij+}^{(5)}(t) + \epsilon \int dt I_{1-}^{ij}(t)I_{0ij+}^{(5)}(t) + {\cal O}(\epsilon^{1.5}) \right].
\label{mqexp}
\end{equation}
The first term corresponds to the 2.5PN radiation reaction, calculated in Appendix~\ref{appendixA}.  The second term can be treated similar to the 2.5PN term because the same factor of $I_{0-}^{ij}$ will be functionally differentiated when applying Eq.~(\ref{eq:vard1}).  In the last term, however, we need to do a bit more work. It is easiest to use Eq.~(\ref{massquad1PN}) and integrate by parts to move the derivatives in the last two pieces of the ``$-$'' mass quadrupole onto the ``$+$'' mass quadrupole in Eq.~(\ref{mqexp}). Once that is done, it is straightforward to vary with respect to the ``$-$'' coordinates following Eq.~(\ref{eq:vard1}). This will be done in Section \ref{sec:rrforce}.

\begin{figure}
	\center
	\includegraphics[width=4cm]{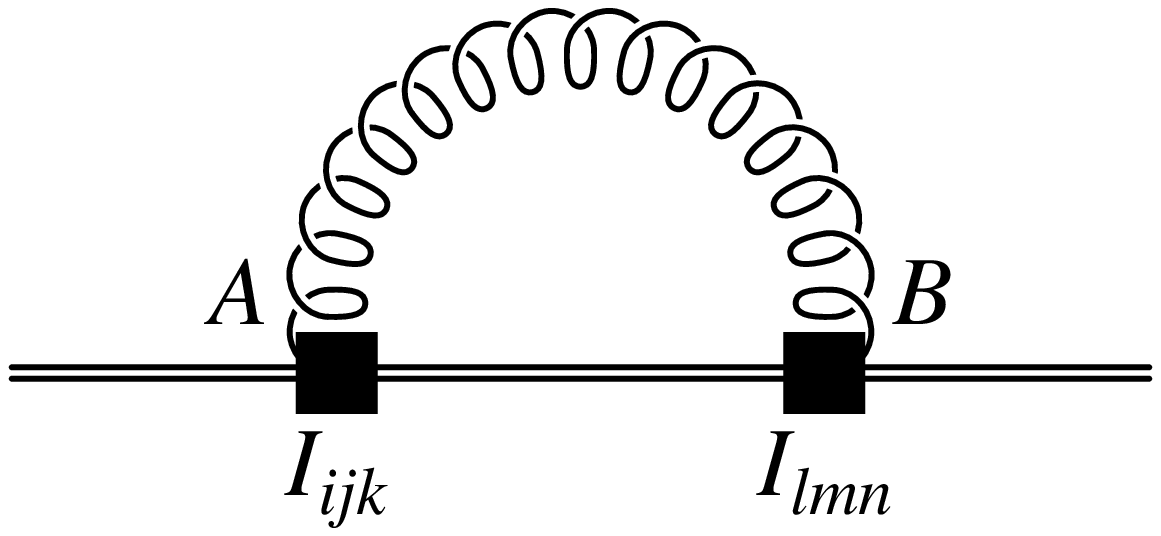}\hspace{1in}\includegraphics[width=4cm]{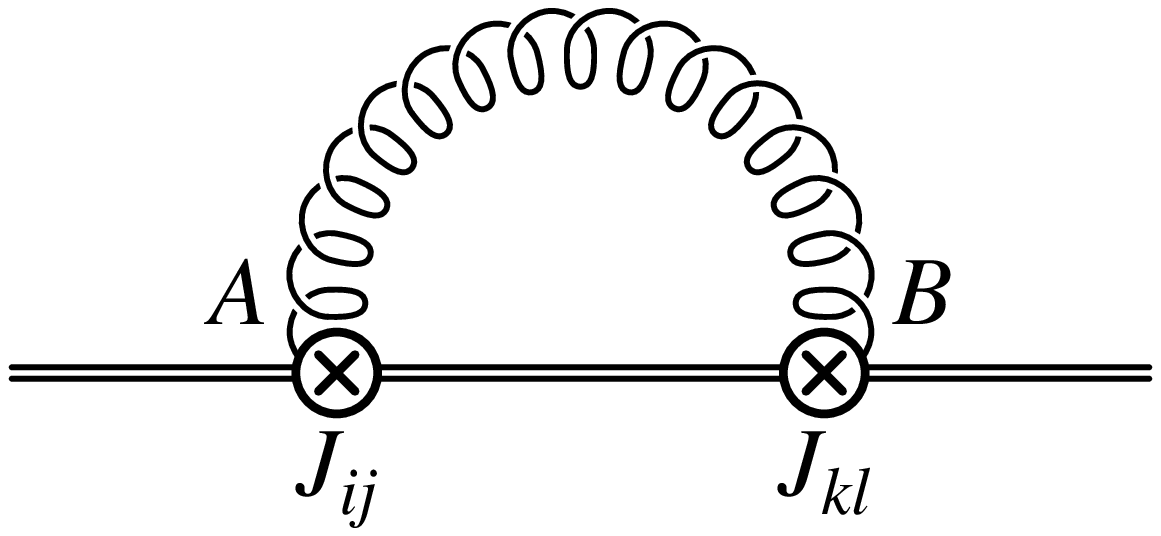}
	\caption{The Feynman diagram giving the contributions to the radiation reaction from the mass octupole (left) and the current quadrupole (right).}
	\label{fig:massoctcurrent}
\end{figure}

There are two more diagrams that need to be calculated in order to get the effective action necessary to extract the 3.5PN radiation reaction force.  First, there is a contribution from the mass octupole, as shown on the left-hand side of Fig.~\ref{fig:massoctcurrent}.  Second, there is a contribution from the current quadrupole shown on the right-hand side of Fig.~\ref{fig:massoctcurrent}.  These will be discussed in turn.

Evaluating the octupole contribution we have
\begin{equation}
i S_{mo} = \frac12 \frac{i^2}{(6m_{pl})^2} \int dt \int dt' \, I^{ijk} _A (t) \big\langle \cE_{ij,k}^A (t) \cE_{lm,n}^B (t') \big\rangle I^{lmn} _B (t'),
\end{equation}
where 
\begin{equation}
\cE_{ij,k} = \frac12\left(h_{00,ijk} + h_{ij,00k} - h_{j0,0ik} - h_{i0,0jk} \right).
\end{equation}
Following the same procedure that is shown in Appendix~\ref{appendixA}, we can write the integrand as
\begin{equation}
I^{ijk} _A (t) \big\langle \cE_{ij,k}^A (t) \cE_{lm,n}^B (t') \big\rangle I^{lmn} _B (t') = I^{ijk} _A (t)  I_{ijkB} (t')\left[\frac52 D^{(3)AB} + 5 \sigma D^{(4)AB} + \sigma^2 D^{(5)AB}\right],
\end{equation}
where $\sigma$ is the Synge world function defined in Eq.~(\ref{eq:sigma1}) and the indices are now contracted with $\delta_{ij}$. 
The quantities $D^{AB}$ are given in Eq.~(\ref{eq:GreenAB}) and the numerical superscripts on $D^{AB}$ indicate the number of derivatives with respect to $\sigma$.
Thus the contribution to the effective action is
\begin{align}
i S_{mo} &= -\frac{1}{72m_{pl}^2} \int dt \int dt'\, I^{ijk} _A (t)  I_{ijkB} (t')\left[\frac52 D^{(3)AB} + 5 \sigma D^{(4)AB} + \sigma^2 D^{(5)AB}\right]\\
&= \frac{8\pi i G}9 \int dt \int dt'\, I^{ijk} _- (t)  I_{ijk+} (t')\left[\frac52 D^{(3)}_{ret} + 5 \sigma D^{(4)}_{ret} + \sigma^2 D^{(5)}_{ret}\right].
\end{align}
Changing variables from $t'$ to $s = t' - t$, holding $t$ fixed,  we find using Eq.~(\ref{eq:sigma2})
\begin{align}
S_{mo} &= \frac{8\pi G}9 \int dt\, I^{ijk} _- (t) \int ds\, 
	\left[-\frac{15}{4s^5}\frac{dD_{ret}(s)}{ds} + \frac{15}{4s^4} \frac{d^2 D_{ret}(s)}{ds^2} 
		- \frac{5}{4s^3}\frac{d^3 D_{ret}(s)}{d3^2} + \frac1{4s}\frac{d^5 D_{ret}(s)}{ds^5}
	\right] I_{ijk+} (t+s).
\end{align}
Integrating by parts to put the derivatives on $I_{ijk+}(t+s)$ and  Laurent expanding  to linear order in $s$, we find
\begin{align}
S_{mo} &= \frac{8\pi G}9 \int dt\, I^{ijk} _- (t) \int ds\, D_{ret}(s)
	\left[\frac{15}{4s^6} I_{ijk+} (t) + \frac{15}{8s^4}  I_{ijk+}^{\prime\prime} (t)
		- \frac{5}{32s^4} I_{ijk+}^{(4)} (t) -\frac{1}{64}  I_{ijk+}^{(6)} (t) - \frac{s}{42} I_{ijk+}^{(7)} (t)
	\right].
\end{align}
The first four terms are divergent and are removed by regularization as described in Appendix \ref{appendixA}.   Keeping only the finite term (i.e., the last one) and using Eq.~(\ref{Dint}), we obtain
\begin{equation}
S_{mo} = -\frac{G}{189}\int dt\, I^{ijk} _- (t) I_{ijk+}^{(7)} (t).
\label{eq:actionMO1}
\end{equation}
Since this contribution begins at order 3.5PN we only need the leading order mass octupole moment $I^{ijk} = I^{ijk}_0 + {\cal O}(\epsilon)$ 
with 
\begin{equation}
I_0^{ijk} = \sum_K m_K \big(x^i_K x^j_K x^k_K\big)_{\rm STF} = -\mu \frac{\delta m}m \big(x^i x^j x^k\big)_{\rm STF},
\end{equation}
where $\delta m = m_1 - m_2$ is the mass difference and STF indicates taking the symmetric, trace-free part of the enclosed expression.
Varying Eq.~(\ref{eq:actionMO1}) using Eq.~(\ref{eq:vard1}) gives the mass octupole contribution to the radiation reaction force at order 3.5PN, the result of which will be given in Section \ref{sec:rrforce}.

The current quadrupole contribution follows similarly.  Evaluating the diagram on the right-hand side of Fig.~\ref{fig:massoctcurrent} we have
\begin{equation}
i S_{cq} =\frac12 \left(\frac{ 2}{ 3 m_{pl} }\right)^2 \int dt \int dt' \, J^{ij} _A (t) \big\langle \cB_{ij}^A (t) \cB_{kl}^B (t') \big\rangle J^{kl} _B (t'),
\end{equation}
where $\cB_{ij}$ is the magnetic part of the Weyl tensor,
\begin{align}
	\cB_{ij} &= \epsilon_{i\alpha \beta \lambda} C^{\alpha \beta} _{~~~ j \rho} u^\lambda u^\rho  \nonumber\\
		&= \epsilon_{i a b} R^{ab} _{~~j0} + O(v),
\end{align}	
and the current quadrupole is given by
\begin{equation}
J^{ij} = \sum_K\bigg(x^i_K L^j_K + \frac32 x^i_K S^j_K\bigg)_{\rm STF},
\end{equation}
where $\vec{L}$ is orbital angular momentum and $\vec{S}$ is the spin angular momentum.  
The spin contribution enters at 4PN so we can neglect the second term in the current quadrupole above.

To linear order, the magnetic Weyl tensor is
\begin{equation}
	\cB_{ij} = \epsilon_{i}^{~ab} \big( h_{a0, bj} + h_{bj, a0} \big).
\end{equation}
Following the same method as discussed in the Appendix, we can write the action as
\begin{align}
i S_{cq} &= - \frac{2 }{ 3 m_{pl}^2} \int dt \int dt' \, J^{ij}_A (t) \big[ \sigma D^{\dprime AB} (\sigma) \big] ^\prime   J_{ijB} (t')\\
&= \frac{ 128\pi i G }{ 3 } \int dt \int dt' \, J^{ij}_- (t) \big[ \sigma D_{ret}^{\dprime} (\sigma) \big] ^\prime J_{ij+} (t') 
\end{align}
Changing variables from $t'$ to $s$, integrating by parts, and Laurent expanding as before we obtain
\begin{equation}
S_{cq} =  - \frac{ 64 \pi G}3  \int dt \, \cJ_- ^{ij} (t) \int ds \, D_{ret} (s) \bigg[ - \frac{ 3}{ s^4} J_{ij+} (t) - \frac{1}{2 s^2} J ^{(2)}_{ij+} (t) + \frac{ 3}{ 8 } J_{ij+} ^{(4)} (t) + \frac{ 4 s}{ 15} J_{ij+}^{(5)} (t) \bigg].
\end{equation}
Again, the first three terms are divergent and are removed by regularization.  Therefore, we obtain 
\begin{equation}
S_{cq} =  - \frac{ 64G}{45} \int dt \, J_- ^{ij} (t) J_{ij+}^{(5)} (t).
\label{eq:actionCQ1}
\end{equation}

\subsection{Radiation reaction force}
\label{sec:rrforce}

We are now ready to vary the action, Eq.~(\ref{Seff}), to obtain the equations of motion.  Since after varying we will set all the ``$-$" variables to zero, we just need to vary with respect to the minus coordinates, using Eq.~(\ref{eq:vard1}), or 
\begin{equation}
F_K^i(t) = \left[\frac{\partial L_{\rm eff}}{\partial x_{Ki-}(t)} - \frac{d}{dt}\left(\frac{\partial L_{\rm eff}}{\partial \dot x_{Ki-}(t)}\right)\right]_{\substack{\vec{x}_{K-}=0 \\ \vec{x}_{K+}=\vec{x}_K }}
\end{equation}
where the effective Lagrangian is given by $S_{\rm eff} = \int dt\, L_{\rm eff}(t)$.  

There are two contributions at order 3.5PN from the mass quadrupole piece, as discussed below Eq.~(\ref{mqexp}).  The easier one is when the lowest order mass quadrupole has the minus coordinates.  In that case, we get
\begin{equation}
a^i_{mq,1} = -\frac{2G}5 x^j I^{(5)}_{1,ij}
\end{equation}
where $x^j$ is the relative coordinate.
To get the other mass quadrupole piece from Eq.~(\ref{mqexp}), first substitute the 1PN part of Eq.~(\ref{massquad1PN}) into the action for the minus coordinates, integrate by parts to remove the explicit derivatives (so that there are no acceleration terms appearing) and then vary with respect to the minus coordinates.  This gives
\begin{align}
a^i_{mq,2} &= \frac{G}{105}(1-3\eta)\left(17 x^i x^j - 11 r^2 \delta^{ij}\right) x^k I^{(7)}_{jk} + \frac{G}{15}(1-3\eta)\left(8x^i x^j v^k - 8 \vec{x} \cdot \vec{v} \, x^k \delta^{ij}  + 9 v^i x^j x^k\right)I^{(6)}_{jk}\nonumber\\
& + \frac{3G}5 (1-3\eta)\left(2 v^i x^j v^k - \frac{m}{r^3} x^i x^j x^k - v^2 x^k \delta^{ij}\right) I^{(5)}_{jk} + \frac{G}5\frac{m}r(1-2\eta)\left(2 x^k\delta^{ij} - \frac1{r^2} x^i x^j x^k\right) I^{(5)}_{jk}.
\end{align}

The mass octupole contribution is, from Eq.~(\ref{eq:actionMO1}),
\begin{equation}
a^i_{mo} = -\frac{G}{63}\mu\frac{\delta m}{m} x^j x^k I^{(7)}_{ijk},
\end{equation}
while the current quadrupole piece is, from Eq.~(\ref{eq:actionCQ1}),
\begin{equation}
a^i_{cq} = \frac{16G}{45} \frac{\delta m}{m}\left( 2 x^j v^k \epsilon^{ikl} J^{(5)}_{jl} + x^k v^j \epsilon^{ikl} J^{(5)}_{jl} - x^j v^k \epsilon^{kjl} J^{(5)}_{il} + x^j x^k \epsilon^{ikl} J^{(6)}_{jl}
\right).
\end{equation}

There is one more place where we get a contribution to the 3.5PN radiation reaction.  We must substitute the 2.5PN radiation reaction for the accelerations that appear on the right hand side of the 1PN acceleration in Eq.~(\ref{1pnnotreduced}).  Doing this gives
\begin{equation}
a^i_{\rm reduced} = \frac{2G}{5}\left[(3+\eta) \frac{m}r x^k \delta^{ij} + \frac{m}{r^3}\eta x^i x^j x^k + \frac12(1-3\eta) v^2 x^k \delta^{ij} + (1-3\eta) v^i x^j v^k\right] I^{(5)}_{jk}. 
\end{equation}

Combining all these terms, the 3.5PN acceleration is 
\begin{align}
a^i_{3.5} &= a^i_{mq,1} + a^i_{mq,2}  + a^i_{mo} + a^i_{cq} + a^i_{\rm reduced} \nonumber\\
&= -\frac{2G}5 x^j I^{(5)}_{1,ij} + \frac{G}{105}(1-3\eta)(17 x^i x^j - 11 r^2 \delta^{ij}) x^k I^{(7)}_{jk} + \frac1{15} (1-3\eta)( 9 v^i x^j + 8 v^j x^i - 8 \vec{x} \cdot \vec{v} \, \delta^{ij})x^k I^{(6)}_{jk} \nonumber\\
&\quad + \frac{G}5\left[ 8(1-3\eta) v^i v^j - (4-13\eta) \frac{m}{r^3} x^i x^j\right] x^k I^{(5)}_{jk} - \frac{2G}5\left[(1-3\eta) v^2 - (4-\eta)\frac{m}r\right] x^k I^{(5)}_{ik}\nonumber\\
& \quad -\frac{G}{63}\frac{\delta m}m x^j x^k I^{(7)}_{ijk} + \frac{16G}{45} \frac{\delta m}{m}\left( 2 x^j v^k \epsilon^{ikl} J^{(5)}_{jl} + x^k v^j \epsilon^{ikl} J^{(5)}_{jl} - x^j v^k \epsilon^{kjl} J^{(5)}_{il} + x^j x^k \epsilon^{ikl} J^{(6)}_{jl}
\right),
\end{align}
where the all the moments are evaluated at lowest order except for $I^{(5)}_{1,ij}$, which is the 1PN contribution.
This agrees with Eq.~(3.15) in Ref.~\cite{IyerWIll:PRD52}. 

\section{Conclusion}
\label{sec:conclusion}

In this paper we computed the radiation reaction forces at the 3.5PN order for a compact binary inspiral using the effective field theory approach. To accomplish this, we utilized a recently introduced extension of Hamilton's variational principle of stationary action \cite{Galley:dissCM} that properly incorporates dissipative effects (e.g., radiation reaction) within a Lagrangian or Hamiltonian formalism. 

We find agreement between our 3.5PN radiation reaction forces and those of Iyer and Will \cite{IyerWill:PRL70, IyerWIll:PRD52}. We also derive the Burke-Thorne 2.5PN radiation reaction force \cite{BurkeThorne:Relativity, Burke:JMathPhys12}, first demonstrated in the EFT approach in \cite{GalleyTiglio:PRD79}, in detail in an appendix. The agreement between our results and previous results gives a strong, non-trivial check that the extended action principle formalism for dissipative mechanics (briefly discussed in Section \ref{sec:dissCM}, its need emphasized in \cite{GalleyTiglio:PRD79}, and given a proper rigorous framework in \cite{Galley:dissCM}) is the correct formalism for describing dissipative effects in a dynamical system.

Combined with previous results from the EFT community, our work nearly completes the equations of motion for a spinning compact binary through 3.5PN order and derived entirely using the EFT approach. These previous works include: the 1PN \cite{GoldbergerRothstein:PRD73, KolSmolkin:CQG25}, 2PN \cite{GilmoreRoss:PRD78}, and 3PN \cite{FoffaSturani:PRD84} non-spinning corrections to the Newtonian potential; the 1.5PN \cite{Porto:PRD73} and 2.5PN \cite{Perrodin:2010dy, Porto:CQG27, Levi:PRD82_1} spin-orbit corrections; the 2PN \cite{Porto:PRD73} and 3PN \cite{Porto:PRD73, PortoRothstein:PRL97, PortoRothstein:0712.2032, PortoRothstein:PRD78} spin-spin corrections; and the 2.5PN \cite{GalleyTiglio:PRD79} and now 3.5PN [this paper] radiation reaction forces. The only contribution remaining to be computed through 3.5PN is from the 3.5PN spin-orbit correction to the potential.

This work also paves the way for higher order radiation reaction calculations in EFT, including the conservative contribution from the radiative back-scattering of gravitational waves off the total mass of the binary spacetime (i.e., the 4PN tail term \cite{BlanchetDamour:PRD37, Blanchet:PRD47, FoffaSturani:2011bq, GL:4PNRR}) and the 4PN contribution from the spin-orbit coupling of the current quadrupole \cite{Wang:2011bt, GL:4PNRR}. 

\section{Acknowledgments}

We thank Ira Rothstein for useful discussions. CRG was supported by an appointment to the NASA Postdoctoral Program at the Jet Propulsion Laboratory administered by Oak Ridge Associated Universities through a contract with NASA. AKL was supported in part by the National Science Foundation under Grant No. PHY-0854782.  Copyright 2012. All rights reserved.

\appendix
\section{Radiation reaction at 2.5PN}\label{appendixA}

To calculate the radiation reaction, we begin by using the worldline action in Eq.~(\ref{radEFT1}) and ``integrate out" the electric and magnetic parts of the Weyl curvature tensor.  This is accomplished by calculating the diagram in Fig.~(\ref{fig:massquad}), where at order 2.5PN the mass quadrupole moment is given in Eq.~(\ref{massquadLO}).  In the dissipative mechanics formulation \cite{Galley:dissCM} of the EFT the contribution to the effective action from this diagram is
\begin{equation}
	i S_{\rm eff}^{\rm 2.5PN} =  \frac12\frac{i^2}{ (2 m_{pl})^2 } \int dt \int dt' \, I^{ij} _A (t) \big\langle \cE_{ij}^A (t) \cE_{kl}^B (t') \big\rangle I^{kl} _B (t')
\label{massquadaction}
\end{equation}
The Feynman rules that tell how to translate Fig.~\ref{fig:massquad} into Eq.~(\ref{massquadaction}) are given in \cite{GalleyTiglio:PRD79}. Here $A$ and $B$ are indices that label the doubled variables (see Sec.~\ref{sec:dissCM}), which will be called {\it history indices}, and the effective action is independent of the basis chosen for the fields. For our purposes, it is convenient to work with the basis in which the history indices are $A,B = \pm$. 

The electric part of the Weyl tensor is given in terms of the metric perturbations to linear order by
\begin{align}
	\cE_{ij} &= R_{\alpha i \beta j} u^\alpha u^\beta  \nonumber\\
		&= \frac12\left[\partial_i \partial_j h_{00}(t,x) + \partial_0 \partial_0 h_{ij}(t,x) - \partial_0 \partial_i h_{j0}(t,x) - \partial_0 \partial_j h_{i0}(t,x) \right] + O(h^2).
\end{align}	
The two-point function in Eq.~(\ref{massquadaction}) is then
\begin{equation}
	\big\langle \cE_{ij}^A (t) \cE_{kl}^B (t') \big\rangle = \frac14  \left\langle[ h^A_{00, ij} (t) + h^A_{ij, 00} (t) ] [ h^B _{00,kl} (t') + h^B_{kl, 00} (t') ] \right\rangle + \frac14  \left\langle[ h^A_{i0, j0} (t) + h^A_{j0, i0} (t) ] [ h^B _{k0,l0} (t') + h^B_{l0, k0} (t') ] \right\rangle,
\end{equation}
where we have evaluated the radiation fields at the binary's center of mass (taken to be at the origin of our spatial coordinates) so that $h_{\alpha \beta} (t) \equiv h_{\alpha \beta} (t, \vec{0})$ and have used
\begin{equation}
	\langle h_{i0}(t) h_{kl}(t') \rangle \propto P_{i0kl} = 0,  
\end{equation}
with
\begin{equation}
P_{\alpha\beta\gamma\delta} = \frac12(\eta_{\alpha\gamma}\eta_{\beta\delta} + \eta_{\alpha\delta}\eta_{\beta\gamma}-\eta_{\alpha\beta}\eta_{\gamma\delta}).
\end{equation}
In terms of the field propagators, the two-point function is
\begin{align}
	\big\langle \cE_{ij}^A (t) \cE_{kl}^B (t') \big\rangle = &\  
		\frac18\left(\partial_i\partial_j \partial_{k'} \partial_{l'} 
			+ 2P_{ijkl} \partial_0^2 \partial_{0'}^2 
			- \eta_{ij} \partial_0^2\partial_{k'}\partial_{l'} 
			- \eta_{kl} \partial_i \partial_j\partial_{0'}^2 \right) D^{AB}(t-t', \vec{0} )\nonumber\\
		& +\frac18\left(\eta_{ik}\partial_j\partial_0 \partial_{l'} \partial_{0'} 
			+ \eta_{jl}\partial_i\partial_0 \partial_{k'} \partial_{0'}
			 + \eta_{il}\partial_j\partial_0 \partial_{k'} \partial_{0'}
			  + \eta_{jk}\partial_i\partial_0 \partial_{l'} \partial_{0'}\right) D^{AB}(t-t', \vec{0} ),
\end{align}
where the prime on the spacetime index of a derivative means that it is a derivative with respect to $x'{}^\mu$ and the matrix of (scalar) propagators in the $\pm$ basis is 
\begin{equation}
D^{AB} = \left(
\begin{array}{cc}
0 & - i D_{adv}\\
-i D_{ret} & 0
\end{array}
\right)  .
\label{eq:GreenAB}
\end{equation}
Since the two-point function is contracted with $I^{ij}_A$ and $I^{kl}_B$, which are symmetric and trace-free, a number of these terms will drop, leaving
\begin{align}
I^{ij}_A(t)\big\langle \cE_{ij}^A (t) \cE_{kl}^B (t') \big\rangle I^{kl}_B(t') = {} &
	\frac18 I_A^{ij}(t)\left[\left(\partial_i\partial_j\partial_{k'}\partial_{l'} 
		+ 2\eta_{ik}\eta_{jl} \partial_0^2\partial_{0'}^2\right) D^{AB}(t-t', \vec{0} )\right] I_B^{kl}(t') \nonumber\\
	&+ \frac12 I_A^{ij}(t)\left[\eta_{ik} \partial_j\partial_0\partial_{l'}\partial_{0'} D^{AB}(t-t', \vec{0} )\right] I_B^{kl}(t')  .
\label{messy}
\end{align}
The propagators can be written in terms of Synge's world function
\begin{equation}
\sigma(x^\alpha , x^{\prime\alpha}) = \frac12 \eta_{\alpha\beta}(x^\alpha - x^{\prime\alpha}) (x^\beta - x^{\prime\beta}) = \frac12(t-t')^2 - \frac12( \vec{x} - \vec{x}{}^{\, \prime} )^2. 
\label{eq:sigma1}
\end{equation}
Therefore, by use of the chain rule, we can simplify Eq.~(\ref{messy}) to
\begin{align}
I^{ij}_A(t)\big\langle \cE_{ij}^A (t) \cE_{kl}^B (t') \big\rangle I^{kl}_B(t') 
	&= I^{ij}_A(t) I_{ijB}(t') \left[\frac32 D^{\prime\prime AB} + 2(t-t')^2 D^{(3)AB} + \frac14(t-t')^4 D^{(4)AB}\right]\\
	&= I^{ij}_A(t) I_{ijB}(t') \left[-\frac12 D^{\prime\prime AB} + \left(\sigma^2 D^{\prime\prime AB}\right)^{\prime\prime} \right],
\end{align}
where the derivatives on $D^{AB}$ are with respect to $\sigma$.  Plugging this back into the effective action, Eq.~(\ref{massquadaction}), we have
\begin{equation}
	i S_{\rm eff}^{\rm 2.5PN} =  -\frac1{ 8 m_{pl}^2 } \int dt \int dt' \, I^{ij} _A (t) I_{ijB} (t')\left[-\frac12 D^{\prime\prime AB} + \left(\sigma^2 D^{\prime\prime AB}\right)^{\prime\prime} \right].
\label{massquadaction2}
\end{equation}
Summing over the history indices $A,B = \pm$ and using Eq.~(\ref{eq:GreenAB}) gives
\begin{align}
iS_{\rm eff}^{\rm 2.5PN} & =  \frac{i}{ 8 m_{pl}^2 } \int dt \int dt' \left\{ I^{ij} _- (t)  I_{ij+} (t')
		\left[-\frac12 D^{\prime\prime}_{ret} + \left(\sigma^2 D^{\prime\prime}_{ret}\right)^{\prime\prime}\right]\right.\nonumber\\
	&\phantom{\frac{i}{ 8 m_{pl}^2 } \int dt \int dt' }\qquad + \left. I^{ij} _+ (t)  I_{ij-} (t')
		\left[-\frac12 D^{\prime\prime}_{adv} + \left(\sigma^2 D^{\prime\prime}_{adv}\right)^{\prime\prime}\right]\right\}\\ 
	& =  8\pi i G \int dt \int dt' I^{ij} _- (t) I_{ij+} (t') 
		\left[-\frac12 D^{\prime\prime}_{ret}(\sigma) + \left(\sigma^2 D^{\prime\prime}_{ret}(\sigma)\right)^{\prime\prime}\right],
\end{align}
where in the last line we have used $m_{pl}^{-2} = 32 \pi G$ and the identity between the retarded and advanced propagators
\begin{equation}
D_{adv}(x,x') = D_{ret}(x',x).
\end{equation}

The next step is to write the integral in the effective action in a quasi-local expansion about the point $t' = t$, which is when the retarded propagator has non-vanishing support.  To do so, define the time difference as
\begin{equation}
s = t'-t,
\end{equation}
so that
\begin{equation}
\sigma = \frac{s^2}2.
\label{eq:sigma2}
\end{equation}
By changing variables to $t$ and $s$ and by integrating by parts to remove the derivatives on the propagators, we can write the effective action as
\begin{align}
S_{\rm eff}^{\rm 2.5PN} &= 8\pi G\int dt I_-^{ij}(t) \int ds\, D_{ret}(s)
	\left[-\frac34\frac{d}{ds}\left(\frac{I_{ij+}(t+s)}{s^3}\right)-\frac34\frac{d^2}{ds^2}\left(\frac{I_{ij+}(t+s)}{s^2}\right)
		-\frac12\frac{d^3}{ds^3}\left(\frac{I_{ij+}(t+s)}{s}\right)\right.\nonumber\\
		&\phantom{8\pi G\int dt I_-^{ij}(t) \int ds\, D_{ret}(s)\qquad}\left.+\frac14\frac{d^4I_{ij+}(t+s) }{ds^4}\right].
\end{align}
The retarded propagator is proportional to 
\begin{equation}
D_{ret}(s, \vec{0} ) \propto \frac{\delta(s)}s,
\end{equation}
so we Laurent expand the terms in the square brackets up $O(s)$, 
\begin{equation}
S_{\rm eff}^{\rm 2.5PN} = 8\pi G\int dt I_-^{ij}(t) \int ds\, D_{ret}(s)
	\left[\frac3{s^4}I_{ij+}(t) +\frac3{8s^2} I_{ij+}^{\prime\prime}(t) + \frac1{32} I_{ij+}^{(4)}(t) + \frac{s}{10} I_{ij+}^{(5)}(t) \right].
\end{equation}

The first three terms are power divergent and are thus zero in dimensional regularization. If we picked a different regularization scheme, these terms will end up renormalizing either the mass or other (possibly gauge-violating) terms in the action (see \cite{GalleyLeibovichRothstein:PRL105} for further discussion on this point).  We thus can remove them and keep only the finite term (the last one),
\begin{align}
S_{\rm eff}^{\rm 2.5PN} &= \frac{4\pi G}5 \int dt I_-^{ij}(t)I_{ij+}^{(5)}(t) \int ds\, s\, D_{ret}(s)\nonumber\\
&= \frac{G}5 \int dt I_-^{ij}(t)I_{ij+}^{(5)}(t),
\label{eq:effaction25c}
\end{align}
where we have used
\begin{equation}
\int ds\,s\,D_{ret}(s) = \frac1{4\pi}.
\label{Dint}
\end{equation}

To get the equations of motion once we have the above action, we vary in the usual way with respect to the worldline coordinates for each body.  At the end of the calculation we set the history indices to be the same so that all the ``$-$" variables are set to zero.  Therefore, to get a non-zero result, we just need to vary with respect to the minus coordinate as in Eq.~(\ref{eq:vard1}), or
\begin{equation}
F_K^i(t) = \left[\frac{\partial L_{\rm eff}^{\rm 2.5PN}}{\partial x_{Ki-}(t)} - \frac{d}{dt}\left(\frac{\partial L_{\rm eff}^{\rm 2.5PN}}{\partial \dot x_{Ki-}(t)}\right)\right]_{ \substack{ \vec{x}_{K-}=0 \\ \vec{x}_{K+} = \vec{x}_K} }
\end{equation}
where the effective Lagrangian is given by $S_{\rm eff}^{\rm 2.5PN} = \int dt\, L_{\rm eff}^{\rm 2.5PN}(t)$.  Using the leading PN mass quadrupole moment, Eq.~(\ref{massquadLO}), we get
\begin{equation}
F_K^i = \frac{2G m_K}5 x_{Kj} \frac{d^5 I^{ij}(t)}{dt}.
\end{equation}
We have been using the mostly-minus convention for the metric, so for the spatial components $\eta^{ij} = -\delta^{ij}$.  Switching now to contracting indices with $\delta^{ij}$ we have
\begin{equation}
F_K^i = -\frac{2G m_K}5 x_{Kj} \frac{d^5 I^{ij}(t)}{dt}.
\end{equation}
We thus recover the Burke-Thorne force \cite{BurkeThorne:Relativity, Burke:JMathPhys12} using the EFT \cite{GalleyTiglio:PRD79}.

\bibliographystyle{iopart-num}
\bibliography{gw_bib}

\setlength{\parskip}{1em}


\end{document}